\documentclass[conference, onecolumn]{IEEEtran}
\IEEEoverridecommandlockouts
\usepackage{cite}
\usepackage{amsmath,amssymb,amsfonts}
\usepackage[utf8]{inputenc} 
\usepackage{booktabs}
\usepackage{algorithmic}
\usepackage{subfig}
\usepackage[normalem]{ulem}
\usepackage{algorithm}
\usepackage{graphicx}
\usepackage{tabularx}
\usepackage{multirow}
\usepackage{textcomp}
\usepackage{xcolor}
\usepackage{url}
\def\BibTeX{{\rm B\kern-.05em{\sc i\kern-.025em b}\kern-.08em
    T\kern-.1667em\lower.7ex\hbox{E}\kern-.125emX}}
\begin{document}

\title{ Enhancing Metaverse Security using Explainable AI-based Intrusion Detection Framework\\
}

\makeatletter
\newcommand{\linebreakand}{%
  \end{@IEEEauthorhalign}
  \hfill\mbox{}\par
  \mbox{}\hfill\begin{@IEEEauthorhalign}
}
\makeatother

\author{
    \IEEEauthorblockN{
        Author 1\IEEEauthorrefmark{1},
        Author 1\IEEEauthorrefmark{1},
        Author 1\IEEEauthorrefmark{1},
        Author 1\IEEEauthorrefmark{1},
        Author 1\IEEEauthorrefmark{1},\\
        Author 1\IEEEauthorrefmark{2}
    }
    
}

\author{\IEEEauthorblockN{Esha Kuppanda\textsuperscript{1}, Shalaka Mahadik\textsuperscript{2}, Pranav M. Pawar\textsuperscript{3}, Mithun Mukherjee\textsuperscript{4}, Raja Muthalagu\textsuperscript{5}}
\IEEEauthorblockA{\textit{Department of Computer Science, BITS Pilani, Dubai Campus, DIAC, UAE \textsuperscript{1,2,3,4,5} } \\
\textit{f20220084@dubai.bits-pilani.ac.in\textsuperscript{1}, mahadikshalaka4@gmail.com\textsuperscript{2},} \\
\textit{pranav@dubai.bits-pilani.ac.in\textsuperscript{3}, mithun@dubai.bits-pilani.ac.in\textsuperscript{4}, raja.m@dubai.bits-pilani.ac.in\textsuperscript{5}}\\
}

}

\maketitle

\begin{abstract}

The MoT virtual environment, called the Metaverse of Things (MoT), combines machine learning (ML) and Internet of Things (IoT) to offer significant business opportunities in various sectors such as entertainment, healthcare, and education. In addition to its benefits, the MoT poses risks to privacy and security through unauthorized access, which threatens trustworthiness. While traditional intrusion detection systems (IDS) exist to handle this challenge, they struggle with high-dimensional data and poor interpretability. To address this challenge, the research conducted a thorough analysis of different ML models and proposed the optimal model to secure the MoT environment. The research employed explainable artificial intelligence (XAI) to identify the most effective features that contributed to optimal model selection. The research presented an optimal framework combining ML and XAI to safeguard MoT networks from the most recent cyberthreats. The research used three advanced benchmark security datasets: ToN-IoT, CIC-IoT-2023, and BoT-IoT, to examine the XAI-based intrusion detection framework. The XAI-based intrusion detection framework achieved 99.83\% accuracy on ToN-IoT and 96.45\% on CIC-IoT-2023 using the XGBoost model. On the BoT-IoT dataset, the Random Forest (RF) and Extra Trees (ET) classifiers reached 99.99\% accuracy. The Shapley Additive Explanations (SHAP) and Local Interpretable Model-Agnostic Explanations (LIME) techniques used for XAI further enhanced feature relevance and user confidence in the optimal model.

\end{abstract}


\section{Introduction}

The MoT integrates machine learning (ML) and Internet of Things (IoT) to enable advanced applications across various domains such as manufacturing, energy, healthcare, education, and many others. As the MoT expands its presence, the attack surface increases, heightening security threats. This endangers user systems by causing operational untrustworthiness alongside data security breaches and privacy violations \cite{CIletter}. The MoT's multi-user and economically driven environment increases vulnerability to cyberattacks, potentially leading to financial losses and reputational damage. The MoT faces a wide range of cyberthreats, including distributed denial of service (DDoS), malware, and ransomware. The complexity of processing data in the virtual environment and detecting these threats surpasses the capabilities of traditional security measures. Many research studies reveal that ML-based intrusion detection systems (IDS) are more prominent in securing the Metaverse and IoT infrastructure \cite{CIletter}. Hence, this study aims to address MoT security using various ML-based techniques.

ML-based IDS offers rapid, adaptive detection with reduced false alarms and can enhance explainability via explainable artificial intelligence (XAI) \cite{globcom2}. However, ML-based IDS faces challenges related to high-dimensional data, real-time adaptation, and interpretability. The nature of applications utilizing IoT devices in a metaverse environment is heterogeneous \cite{ssm2}. They generate large volumes of data that need to be processed securely. Therefore, the research performed an extensive comparative analysis using various ML techniques and proposed an optimal model to secure the MoT environment from various threats. The research also included XAI techniques, namely, Shapley Additive Explanations (SHAP) and Local Interpretable Model-Agnostic Explanations (LIME); these techniques helped to show which features were more influential and assisted in selecting the optimal model. The research used ToN-IoT \cite{b1}, CIC-IoT-2023 \cite{cic-dataset}, and BoT-IoT \cite{bot-dataset} datasets to simulate diverse cyberthreats. The proposed XAI-based intrusion detection model is not only accurate and transparent, but it also ensures end-user trust and a safe ecosystem for MoT applications. End-user trust is of paramount importance in today's MoT domain.

\noindent The main contributions of the research are as follows:

\begin{itemize}

\item The research undertook an in-depth analysis of numerous ML-based models, including RF, Decision Tree (DT), AdaBoost, ET, XGBoost, Multi-layer Perceptron (MLP), and Gradient Boosting (GB). The research focused on thorough feature engineering and hyperparameter optimization.

\item The research employed XAI techniques such as SHAP and LIME to provide an accurate explanation of the model's most influential features. The XAI's transparent and trustworthy decision helps to gain end-users' confidence.


\item The research developed an XAI-based intrusion detection framework for threat detection (binary) or attack classification, i.e., attack category or subcategory in MoT environments. The proposed framework enables real-time monitoring, alerts, and security decision-making.  

\item The research utilized three benchmark IoT security datasets, namely, BoT-IoT, ToN-IoT, and CIC-IoT-2023, for performance analysis.

\item The proposed XAI-based intrusion detection framework outperforms existing IDS models. The research considered four standard performance metrics, namely, F1-score, recall, precision, and accuracy, to assess and compare the proposed framework.

\end{itemize}

\noindent The organization of the paper is as follows: related work is discussed in Section 2, the Metaverse security framework is discussed in Section 3, and the architecture, data processing, and ML techniques employed are discussed in Section 4. The results and discussion are presented in Section 5. The conclusions and future scope are presented in Section 6.

\section{Related work}

The section explores various state-of-the-art works conducted to address Metaverse security threats and XAI methods for enhancing trustworthiness. The paper \cite{b1} presents a distributed edge-based IDS tailored to IoT frameworks, capable of rapidly detecting intrusions at the network's edge. A hybrid DL-based IDS presented in \cite{b2} detects multiple attack types and enhances detection rates through DL model training. The paper \cite{b3} proposes an anomaly detection system that combines feature selection with deep learning to enhance intrusion detection capability in IoT networks. The paper \cite{b5} presents a hybrid DL-based IDS with a scalable architecture designed to analyze real-time IoT traffic. The study \cite{b6} presents an IDS using various ML techniques, whereas the paper \cite{b7} presents real-time anomaly detection based on sequential learning methods; both attained good accuracy.

Next, the paper \cite{b8} presents an IDS based on feature reduction to enhance detection efficiency in IoT environments. The study \cite{b9} combines game theory, transfer learning, and big data processing to accelerate IDS decision-making at the cost of high computational complexity. A CNN-LSTM-based IDS presented in \cite{b10} successfully captures both spatial and temporal attack patterns but incurs significant computational overhead. The paper \cite{b11} focuses on advanced feature engineering to enhance the performance of IDS. The study \cite{b12} tests standard ML models for IoT intrusion detection, demonstrating acceptable performance. The paper \cite{b13} presents a hybrid DL-based IDS that achieves good detection accuracy, but its computational costs make real-time implementation infeasible. The paper \cite{b14} presents a transformer-based IDS that improves multi-class detection accuracy but consumes substantial memory resources. Lastly, the paper \cite{b15} proposes a stacked ensemble model that has better resilience against known and unknown attacks but is not readily deployable on lightweight IoT devices.

\begin{table*}[ht]

\centering

\caption{The state-of-the-art Related work using ML techniques} \label{RWtt}

\scriptsize

\setlength{\tabcolsep}{3pt}

\renewcommand{\arraystretch}{1.1}


\begin{tabular}{c p{4.5cm} p{3cm} p{3cm} p{2.5cm} c}

\toprule


\textbf{Year} & \textbf{Main idea and ref no.} & \textbf{ML Techniques} & \textbf{XAI Method} & \textbf{Dataset} & \textbf{Accuracy(\%)} \\

\midrule

2021 & AI-driven IDS for IoT security at the network edge with real-time detection \cite{b1}. & LSTM, RNNs & No explicit XAI & ToN-IoT & 99.89 \\

2022 & Enhance IDS reliability for DDoS, DoS, Brute force, and Botnet attacks \cite{b2}. & DL techniques & Feature selection & BoT-IoT & 98.90 \\

2022 & Deep learning IDS for smart home IoT with real-time detection \cite{b3} & PCC-CNN model & No XAI & BoT-IoT & 92.85 \\


2023 & Real-time IDS for large-scale IoT using ensemble deep learning \cite{b5} & LSTM, CNN, ANN & No XAI & ToN-IoT & 99.93 \\

2023 & IDS design for detecting multiple IoT attacks \cite{b6} & ML techniques & Feature selection & BoT-IoT & 99.8 \\

2023 & Real-time anomaly detection using sequential patterns \cite{b7} & LSTM model & No XAI & CIC-IoT-2023 & 98 \\

2024 & Metaverse IoT IDS using rough set-based feature selection \cite{b8} & ML techniques & Rough set theory & ToN-IoT, CIC-IoT-2023 & 99.7 \\

2024 & IoT IDS using transfer learning and game theory \cite{b9} & DL techniques & Game Theory & CIC-IoT-2023 & 98.2 \\

2024 & CNN-LSTM based IDS for IoT security \cite{b10} & CNN, LSTM & No XAI & CIC-IoT-2023 & 98.42 \\

2024 & Improve ML-based IDS efficiency with feature engineering \cite{b11} & ML techniques & No XAI & BoT-IoT & 98.49 \\

2024 & Evaluate IDS performance on ToN-IoT Linux datasets \cite{b12} & DT, RF & No XAI & ToN-IoT & 95 \\

2024 & Hybrid CNN-LSTM IDS for improved detection \cite{b13} & DL techniques & No XAI & CIC-IoT-2023, ToN-IoT & 98-99 \\

2024 & Transformer-based multi-class IDS for  \cite{b14} & CNN, LSTM models & No XAI & CIC-IoT-2023 & 99.4 \\

2024 & Botnet detection, including zero-day attacks using multi-classifier \cite{b15} & ML approach & No XAI & BoT-IoT & 98 \\

\multicolumn{2}{l}{\textbf{XAI-based intrusion detection framework for MoT}} & \textbf{RF, DT, ET, GB, MLP, AdaBoost, XGBooST} & \textbf{XAI: SHAP, LIME} & \textbf{ToN-IoT,CIC-IoT-2023, BoT-IoT} & \textbf{99.83, 96.45, 99.99} \\

\bottomrule

\end{tabular}

\end{table*}

Table \ref {RWtt} summarizes the state-of-the-art related work using various ML techniques, XAI methods, datasets, and performance metrics. 
Table \ref{RWtt} highlights that various IDS have been developed to address IoT security challenges. The ML-based IDS approaches reviewed in the literature face several limitations, including high computational requirements (particularly for hybrid models), real-time detection latency, and limited explainability arising from their black-box nature. Furthermore, existing studies primarily focus on conventional IoT security and give limited consideration to the distinct security challenges arising from the integration of IoT devices within Metaverse environments.

Motivated by these limitations, this research aims to develop a unified IDS framework that achieves high detection accuracy and a low error rate while providing transparent and interpretable predictions. Accordingly, the research proposes an XAI-based intrusion detection framework for MoT security. The framework employs SHAP and LIME to explain model predictions, identify influential features, and improve user confidence in the intrusion detection process.

\section{MoT security using XAI-based intrusion detection}

The section outlines MoT security using an XAI-based intrusion detection framework. Fig.\ref{archi-fig} details the pipeline of the proposed XAI-based intrusion detection for the MoT ecosystem.

\begin{figure}[htbp] 

\centerline{\includegraphics[width=\columnwidth]{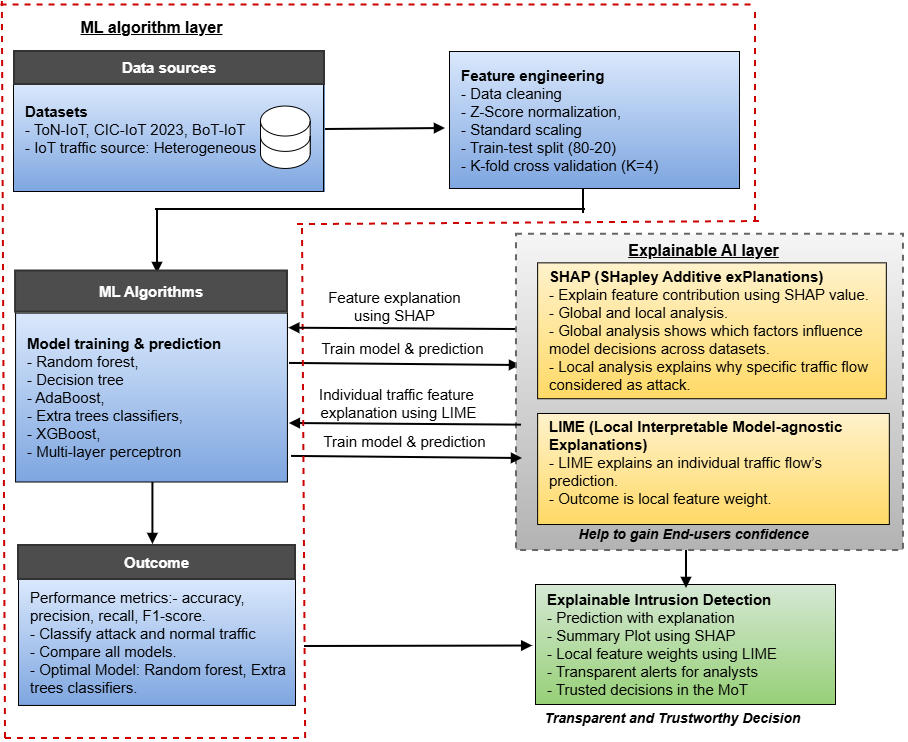}} 

\caption{Block diagram for proposed XAI-based intrusion detection framework.} \label{archi-fig} 

\end{figure}

As shown in Fig.\ref{archi-fig}, Metaverse virtual environments connect IoT devices in a complex network, exposing them to various cyberthreats. The research used three benchmark IoT security datasets: BoT-IoT, ToN-IoT, and CIC-IoT-2023—to develop a robust IDS, as they give distinct insights into different types of attacks and network traffic patterns. The research first cleaned and processed the raw data to improve its quality. Missing data was addressed using statistical or model-based imputation, with categorical attributes converted to numerical values and features normalized to ensure consistent scaling.

Moreover, the SelectKBest using ANOVA technique was employed to select important features and to reduce model overfitting \cite{anova}. The research also utilized four-fold cross-validation to facilitate a reliable performance evaluation \cite{kfold}. 

The research used RF, DT, XGBoost, AdaBoost, ET, GB, and MLP models to train and predict the outcome. The XAI layer with SHAP assists the research in clarifying model predictions and various influential features. Similarly, LIME examines each traffic flow and defines the features that aid in understanding the predicted knowledge. Finally, the ML and XAI layers help to select the optimal model and enable real-time monitoring, warnings, and security decision-making.

\section{Methodology}

This section describes the comprehensive data preparation and ML approaches applied to three IoT datasets. During the feature engineering stage, we addressed missing value imputation, categorical encoding, feature scaling, and feature selection to enhance model performance. The datasets were standardized and split into training and testing sets. ML models such as RF, DT, ET, GB, XGBoost, AdaBoost, and MLP were used for predictive analysis and evaluated using accuracy, classification report, and confusion matrix. Feature engineering helped achieve reliable evaluation, and the top models for each dataset were chosen based on their optimal accuracy \cite{ssm2}.

\subsubsection*{ToN-IoT data preparation:}
 
The pandas package was used to load data from a CSV file containing features and labels. Missing values were then handled as follows: for numerical columns, missing values were replaced with the mean, while for categorical columns, missing values were replaced with the mode. The preprocessed data was split into training and testing sets.

\subsubsection*{BoT-IoT data preparation:}
Due to its large size, the BoT-IoT dataset was preprocessed using a pipeline that includes chunk-based reading, cleaning, encoding, scaling, and preparation for ML. In terms of data cleaning, empty strings and invalid values, such as infinities, were replaced with NaN values, while missing data was handled by imputing numerical features with their means and categorical features with their modes or placeholder values. Target and other categorical features were encoded using LabelEncoder to maintain consistent label representation across categories, while numerical features were standardized using StandardScaler to prepare the data for model training. The cleaned and normalized data was further split into training and testing sets using stratified splitting to preserve class distribution.

\subsubsection*{CIC-IoT-2023 data preparation:}
For the CIC-IoT-2023 dataset, the Python script first reads a CSV file containing network-flow features and verifies the presence of all required columns, including the target column, \texttt{label}. It then separates the features, denoted by $X$, from the target vector, denoted by $y$. The \texttt{LabelEncoder} method converts the categorical target labels into numerical values. Finally, the dataset is divided into training and testing sets using a stratified train-test split to preserve the original class distribution.  

\subsubsection*{ML model training, testing, and prediction:}

Following data preparation, the research investigated several ML algorithms for training and threat detection in the MoT environment, including RF, DT, ET, GB, XGBoost, AdaBoost, and MLP. Each model was trained on the training data and evaluated on the test data. The model with the highest accuracy score was considered the optimal one. To explain which features contribute to the optimal model's performance, XAI techniques such as SHAP were used across all ML models to calculate global and instance-level feature importance and individual feature contributions. LIME, another XAI approach, provided local feature importance by approximating the model's behavior using perturbed samples. When assessing Random Forest's built-in feature importance, permutation importance was utilized to calculate the decrease in model accuracy when feature values are randomly permuted. These models were then evaluated using accuracy, precision, recall, and F1-score.
Table \ref{dataset-detail} presents dataset details and the number of selected features. The hyperparameter settings are outlined in Table \ref{hype-para}.

\begin{table}[htbp]
  \centering
  \caption{Dataset details and selected features}
  \label{dataset-detail}
  \scriptsize
  \setlength{\tabcolsep}{2pt}
  \renewcommand{\arraystretch}{1.0}
  \resizebox{\columnwidth}{!}{%
    \begin{tabular}{l r r r c}
    \toprule
    Dataset & Total samples & Training samples & Testing samples & Features \\
    \midrule
    ToN-IoT & 21,104 & 16,883 & 4,221 & 15 \\
    \midrule
    CIC-IoT-2023 & 1,279,753 & 1,023,802 & 255,951 & 46 \\
    \midrule
    BoT-IoT & 733,705 & 586,964 & 146,741 & 7 \\
    \bottomrule
    \end{tabular}%
  }
\end{table}%

\begin{table}[htbp]
  \centering
  \caption{Hyperparameter settings for ML models}
  \label{hype-para}
  \scriptsize
  \setlength{\tabcolsep}{2pt}
  \renewcommand{\arraystretch}{1.08}
  \resizebox{\columnwidth}{!}{%
    \begin{tabular}{|l|p{0.70\columnwidth}|}
    \toprule
    ML models & Hyperparameters \\
    \midrule
    \multicolumn{2}{|c|}{ToN-IoT} \\
    \midrule
    RF & n\_estimators = 100, criterion = gini \\
\cmidrule{1-2}    AdaBoost & n\_estimators = 50, learning\_rate = 1.0, algorithm = SAMME \\
\cmidrule{1-2}    XGBoost & n\_estimators = 100, learning\_rate = 0.3, max\_depth = 6, eval\_metric = logloss \\
\cmidrule{1-2}    ET & n\_estimators = 100, criterion = gini \\
\cmidrule{1-2}    DT & criterion = gini, splitter = best, random\_state = None \\
\cmidrule{1-2}    GB & loss = log\_loss, learning\_rate = 0.1, n\_estimators = 100 \\
\cmidrule{1-2}    MLP & hidden\_layer\_sizes = (100,), activation = relu, solver = adam, max\_iter=300 \\
    \midrule
    \multicolumn{2}{|c|}{BoT-IoT} \\
    \midrule
    RF & class\_weight = balanced, n\_estimators = 50, criterion = gini \\
\cmidrule{1-2}    AdaBoost & n\_estimators = 50, learning\_rate = 1.0, algorithm = SAMME \\
\cmidrule{1-2}    XGBoost & n\_estimators = 50, learning\_rate = 0.3, eval\_metric = logloss \\
\cmidrule{1-2}    ET & class\_weight = balanced, n\_estimators = 50, criterion = gini \\
\cmidrule{1-2}    DT & class\_weight = balanced, criterion = gini \\
\cmidrule{1-2}    MLP & hidden\_layer\_sizes = (100,), activation = relu, solver = adam, max\_iter = 300 \\
    \midrule
    \multicolumn{2}{|c|}{CIC\_IoT-2023} \\
    \midrule
    RF & n\_estimators = 100, criterion = gini, random\_state = 42 \\
\cmidrule{1-2}    AdaBoost & n\_estimators = 50, algorithm = SAMME, random\_state = 42 \\
\cmidrule{1-2}    XGBoost & n\_estimators = 100, eval\_metric = mlogloss \\
\cmidrule{1-2}    ET & n\_estimators = 100, criterion = gini, random\_state = 42 \\
\cmidrule{1-2}    DT & criterion = gini, random\_state = 42 \\
\cmidrule{1-2}    MLP & hidden\_layer\_sizes = (100,), activation = relu, solver = adam, max\_iter = 300, random\_state = 42 \\
    \bottomrule
    \end{tabular}%
 }
\end{table}%


\section{RESULTS AND DISCUSSIONS}
\subsection{Performance evaluation using ToN-IoT dataset:}
Table \ref{tt-ton1} shows that most ML models performed well, with RF, XGBoost, ET, GB, and MLP being particularly effective. The leading algorithm, XGBoost, achieved the highest accuracy score of 99.83\%.

\begin{table}[h]
\centering
\caption{Performance evaluation using ToN-IoT dataset}
\label{tt-ton1}
\footnotesize
\renewcommand{\arraystretch}{1.1}
\setlength{\tabcolsep}{3.5pt}
\begin{tabular}{|l|c|c|c|c|}
\hline
\textbf{Model} & \textbf{Acc.} & \textbf{Prec.} & \textbf{Rec.} & \textbf{F1} \\ \hline
RF & 99.74 & 100.0 & 100.0 & 100.0 \\ \hline
AdaBoost & 99.38 & 99.0 & 99.0 & 99.9 \\ \hline
\textbf{XGBoost} & \textbf{99.83} & 100.0 & 100.0 & 100.0 \\ \hline
ET & 99.79 & 100.0 & 100.0 & 100.0 \\ \hline
DT & 99.36 & 99.0 & 99.0 & 99.0 \\ \hline
GB & 99.79 & 100.0 & 100.0 & 100.0 \\ \hline
MLP & 99.67 & 100.0 & 100.0 & 100.0 \\ \hline
\end{tabular}
\end{table}

\subsubsection{ToN-IoT Feature Importance}
Fig. \ref{img-21} and Fig. \ref{img-22} present the permutation importance and SHAP summary plots for the MLP classifier. The plots quantify each feature's contribution and show how permuting each feature affects the model performance. Features with relatively low importance include \path{Memory_pct_Committed_Bytes_In_Use}, \path{LogicalDisk(_Total)_Disk_Read_Bytes_sec}, and \path{Memory_Available_Bytes}. The low importance of \path{Processor_DPC_Rate} indicates that it contributes little to the classifier's predictions, whereas network-related features have greater influence. Together, permutation importance, SHAP, and LIME provide a reliable and interpretable assessment of feature relevance for the dataset.

\begin{figure}[htbp]
    \centering
    \includegraphics[width=0.70\linewidth]{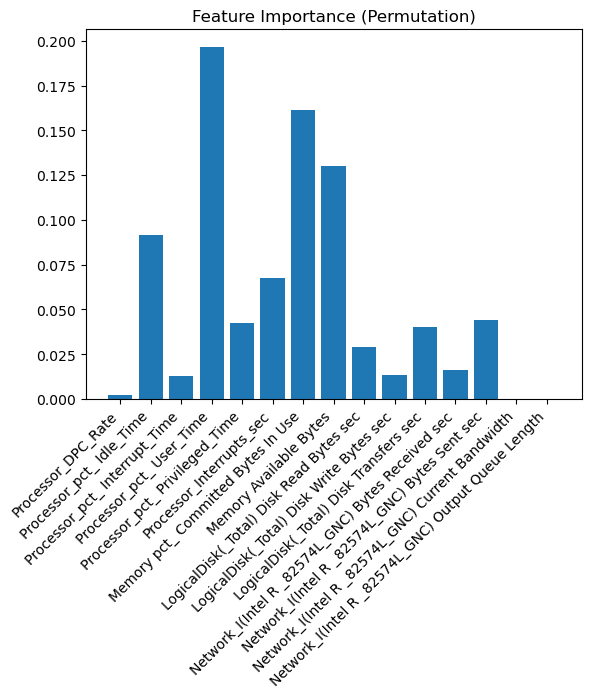}
    \caption{Feature Importance using XAI for ToN-IoT dataset}
    \label{img-21}
\end{figure}

\begin{figure}[htbp]
    \centering
     \includegraphics[width=0.70\linewidth]{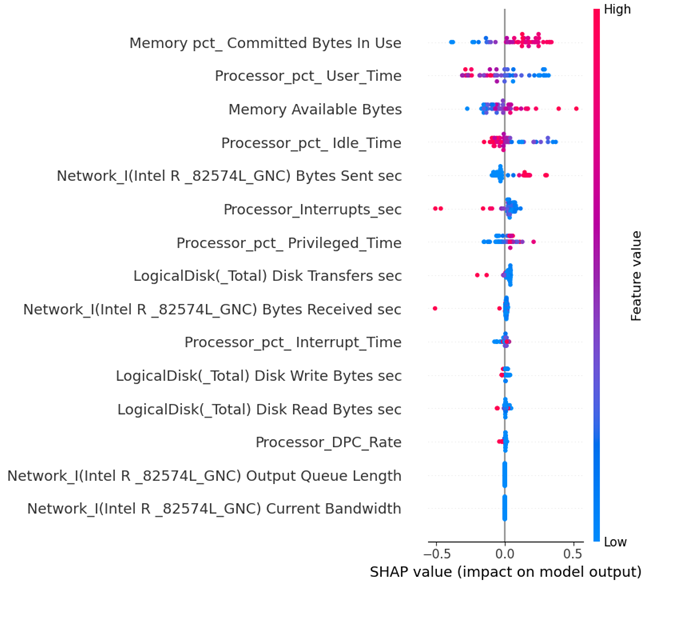}
    \caption{Summary Plot using XAI for ToN-IoT dataset}
    \label{img-22}
\end{figure}

\subsection{Performance evaluation using CIC-IoT-2023 dataset:}
Table \ref{tt-cic1} compares the accuracy, precision, recall, and F1-score of various ML-based models. XGBoost is the best-performing model, achieving 96.45\% accuracy with similarly high precision, recall, and F1-score, followed closely by RF with 96.37\% accuracy. Furthermore, the ET and DT models show satisfactory accuracies of 95.57\% and 95.70\%, demonstrating reliable predictive performance. In summary, XGBoost is the top-performing model, with the best precision and strong performance across all metrics used in the evaluation.


\begin{table}[h]
\centering
\caption{Performance evaluation using CIC-IoT-2023 dataset.}
\label{tt-cic1}
\footnotesize
\renewcommand{\arraystretch}{1.1}
\setlength{\tabcolsep}{4pt}
\begin{tabular}{|l|c|c|c|c|}
\hline
\textbf{Model} & \textbf{Acc.} & \textbf{Prec.} & \textbf{Rec.} & \textbf{F1} \\ \hline
RF   & 96.37 & 93.22 & 85.92 & 88.80 \\ \hline
AdaBoost  & 70.47 & 50.12 & 39.78 & 38.11 \\ \hline
\textbf{XGBoost}  & \textbf{96.45} & 91.95 & 87.09 & 89.15 \\ \hline
ET    & 95.57 & 91.93 & 83.13 & 86.39 \\ \hline
DT   & 95.70 & 86.83 & 86.96 & 86.90 \\ \hline
MLP  & 47.78 & 52.99 & 16.50 & 13.48 \\ \hline
\end{tabular}
\end{table}

\subsubsection{XAI Feature Importance for CIC-IoT-2023}
The SHAP technique was used to construct the feature importance chart, which shows the degree to which each feature contributes to the MLPClassifier model's predictions. It also shows that features with higher mean SHAP values, such as \path{flow_duration}, \path{rst_count}, and \path{Min}, have a greater impact on the model's predictions. The technique involved normalizing the dataset before producing SHAP values from a sample of the training data, which were used to assess feature importance. Although the SHAP values were approximated from a sample of the dataset, Fig. \ref{img41} and Fig. \ref{img42} show the relative magnitude and impact of each feature on the model in descending order. This enables the identification of key features that most significantly contribute to predictive performance, thereby aiding model interpretability, optimization, and explanation.

\begin{figure}[htbp]
    \centering
    \includegraphics[width=0.85\linewidth]{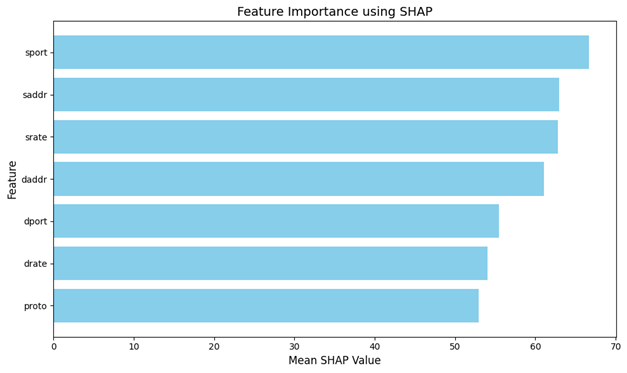}
    \caption{Feature Importance using XAI for CIC-IoT-2023 dataset}
    \label{img41}
\end{figure}

\begin{figure}[htbp]
    \centering
    \includegraphics[width=0.85\linewidth]{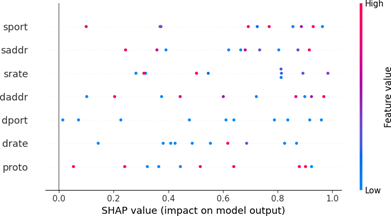}
    \caption{SHAP Summary Plot for CIC-IoT-2023 dataset}
    
    \label{img42}
\end{figure}

\subsection{Performance evaluation using BoT-IoT dataset:}

Table \ref{tt-bot} compares the performance of different ML algorithms in terms of attack, category, and subcategory accuracy, as well as precision, recall, and F1-score. RF and ET classifiers achieved the highest performance, both attaining 99.99\% attack and category accuracy. The subcategory accuracy was 99.98\% for RF and 99.97\% for ET. Furthermore, both achieved 100\% precision, recall, and F1-score. Other models, including XGBoost and DT, also performed well.

\begin{table}[h]
\centering
\caption{Performance evaluation using BoT-IoT dataset}
\label{tt-bot}
\footnotesize
\renewcommand{\arraystretch}{1.1}
\resizebox{0.5\textwidth}{!}{%
\begin{tabular}{|l|c|c|c|c|c|c|}
\hline
\textbf{Model} & \textbf{AttackAcc} & \textbf{CategoryAcc} & \textbf{SubcategoryAcc} & \textbf{Prec} & \textbf{Rec} & \textbf{F1} \\ \hline
\textbf{RF}   & \textbf{99.99}  & 99.99 & 99.98 & 100.0 & 100.0 & 100.0 \\ \hline
AdaBoost  & 95.88  & 88.67 & 98.99 & 91.25 & 99.00 & 94.02 \\ \hline
XGBoost  & 99.86  & 99.60 & 99.97 & 99.57 & 99.61 & 99.59 \\ \hline
\textbf{ET}   & \textbf{99.99}  & 99.99 & 99.97 & 100.0 & 100.0 & 100.0 \\ \hline
DT   & 99.98  & 99.98 & 99.98 & 100.0 & 100.0 & 100.0 \\ \hline
MLP  & 97.39  & 92.46 & 99.71 & 97.51 & 99.64 & 99.10 \\ \hline

\end{tabular}
}
\end{table}

\subsubsection{XAI Feature Importance for BoT-IoT}

Fig. \ref{img71} and Fig. \ref{img72} show the relative influence of seven features on the MLP classifier's prediction of the \texttt{attack} target. The SHAP results indicate that \path{sport}, \path{saddr}, and \path{srate} have the greatest influence on the model's predictions, followed by \path{daddr} and \path{dport}. Although \path{drate} and \path{proto} have lower SHAP values, they still provide useful predictive information. Overall, the plots identify the features that drive the model's predictions and provide a clearer understanding of its behavior.

\begin{figure}[htbp]
    \centering
    \includegraphics[width=0.85\linewidth]{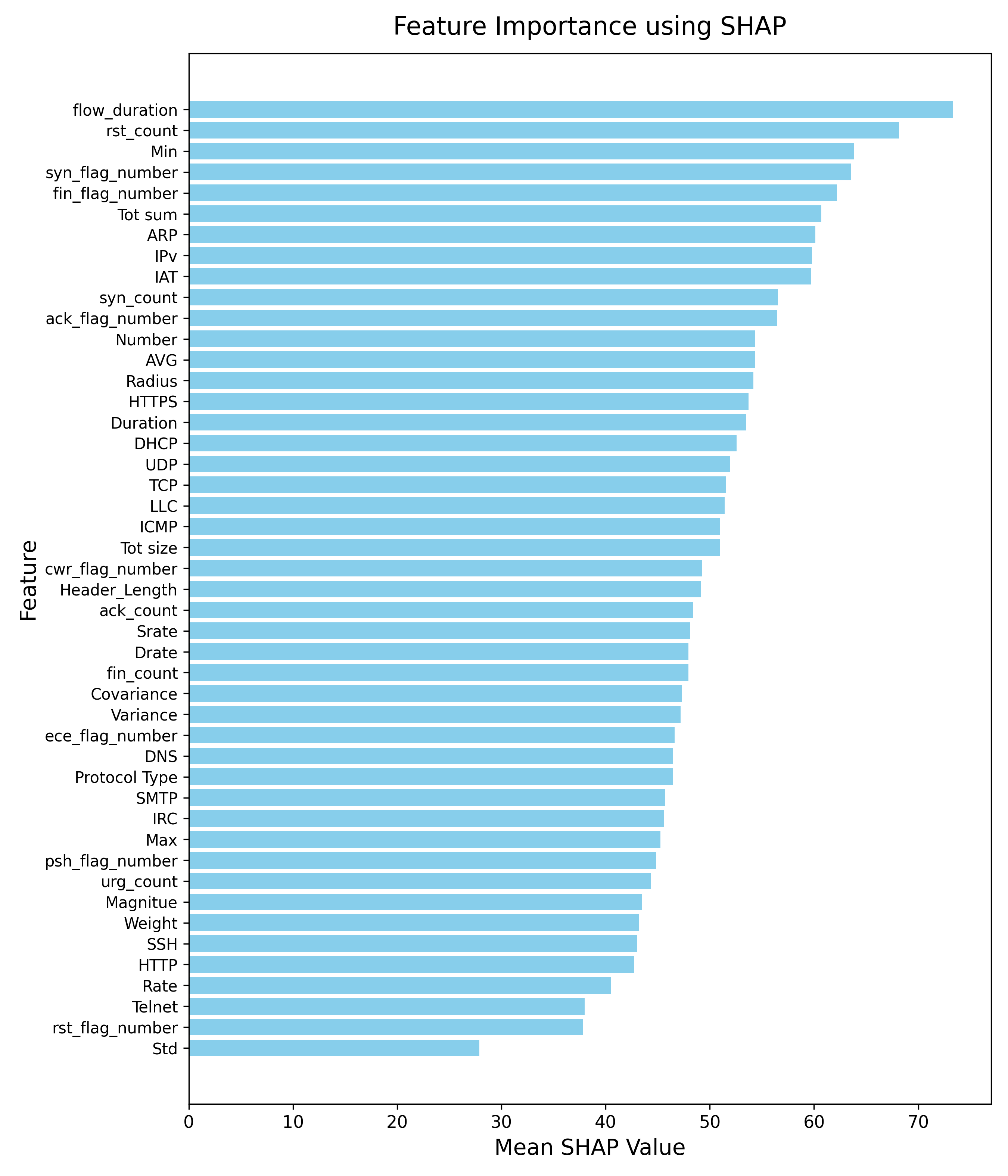}
    \caption{Feature Importance using XAI for BoT-IoT}
    \label{img71}
\end{figure}

\vspace{-5pt}

\begin{figure}[htbp]
    \centering
    \includegraphics[width=0.85\linewidth]{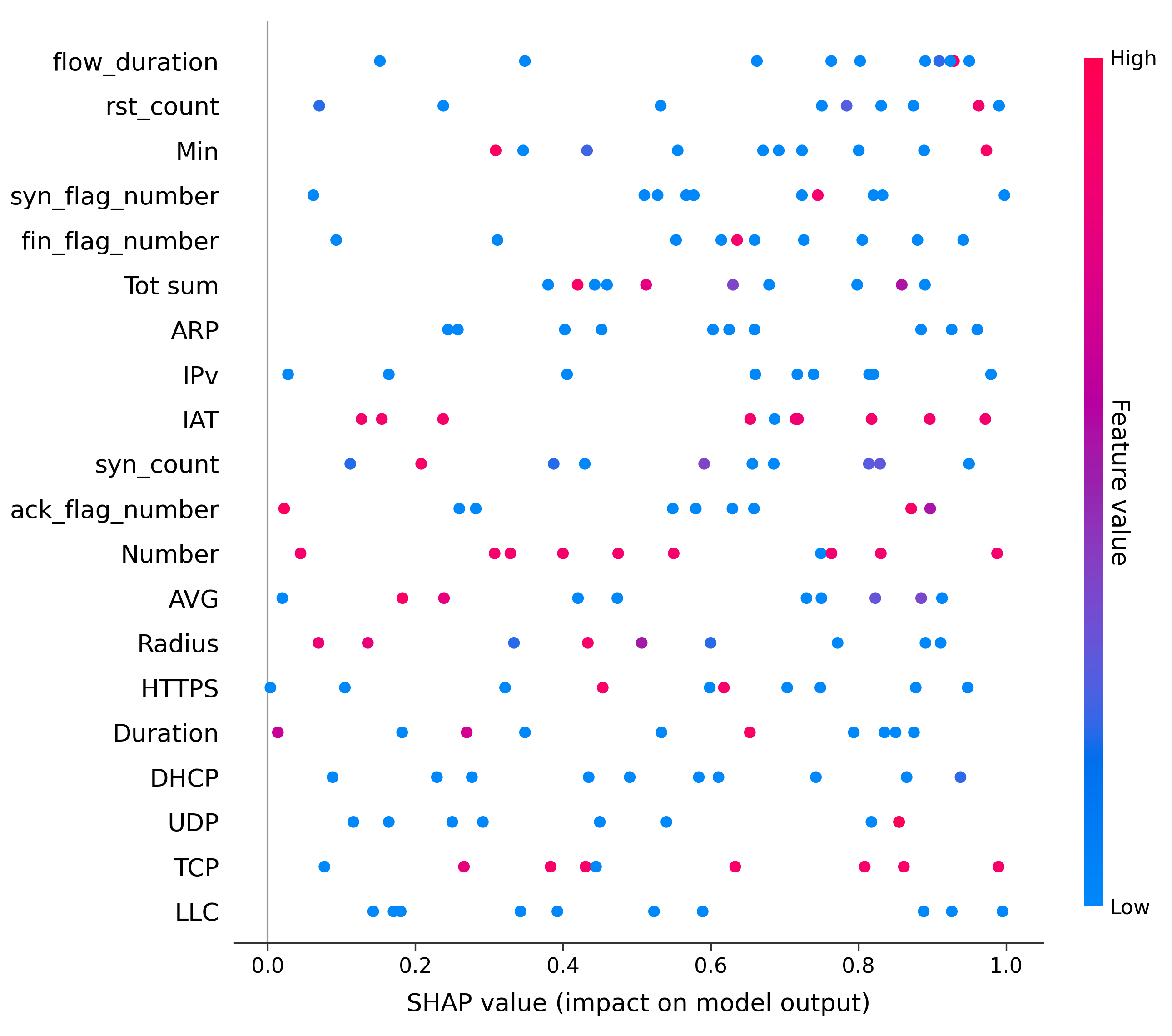}
    \caption{SHAP Summary Plot for BoT-IoT}
    \label{img72}
\end{figure}

\section{CONCLUSION}

Metaverse virtual environments connect IoT devices in a complex network, exposing them to a variety of cyber threats. Given the numerous possibilities for cybercrime in metaverse environments, a single detection approach is insufficient to provide comprehensive protection against all threats.
To address this concern, this study proposes an XAI-based intrusion detection framework to safeguard the Metaverse of Things (MoT) environment against diverse cyber threats. A comprehensive analysis of standard ML models — namely RF, DT, ET, GB, XGBoost, AdaBoost, and MLP — was conducted, encompassing feature engineering, the training and testing phase, and XAI integration.
XAI techniques, specifically SHAP and LIME, were employed to identify and interpret critical features in MoT scenarios; for instance, which network-related features are most vulnerable when IoT devices communicate within the metaverse environment. XAI promotes transparency and user trust by enabling users to understand the reasoning employed by ML models to arrive at a given prediction, thereby increasing confidence in the ML-based IDS.

The proposed framework demonstrated notable efficiency and accuracy, leveraging extensive feature engineering for feature extraction from three recent benchmark datasets. It achieved 99.83\% accuracy on ToN-IoT and 96.45\% on CIC-IoT-2023 using the XGBoost model, while RF and ET achieved 99.99\% accuracy on the BoT-IoT dataset. Hyperparameter tuning and feature engineering enabled the framework to attain maximum accuracy across all three benchmark IoT datasets. The SHAP and LIME techniques enhanced the interpretability of MoT-based detection and increased user confidence in the system's operation. Evaluation results confirmed that the framework performs effectively in real-time security testing of decentralized complex systems.
Future work could extend this research by incorporating deep learning to further strengthen security in MoT environments.

\bibliographystyle{IEEEtran}
\bibliography{references}

\end{document}